\documentclass[a4paper,12pt]{article}
 
\pdfoutput=1
\usepackage{jheppub}
\usepackage{dcolumn}
\usepackage{soul}
\usepackage[english]{babel}
\usepackage[utf8]{inputenc}
\usepackage{dcolumn}
\usepackage{soul}
\usepackage{hyperref}
\usepackage{url}
\usepackage{graphicx}
\usepackage{bm,amsmath,amssymb}
\usepackage[mathscr]{eucal}
\usepackage{makeidx}
\usepackage{subfig}
\usepackage{amsmath}
\usepackage{amssymb}
\usepackage{amsthm}
\usepackage{mathrsfs}
\usepackage{graphicx}
\usepackage{fancyhdr}
\usepackage{array}
\usepackage{simplewick}
\usepackage{latexsym}
\usepackage[all]{xy}
\usepackage{enumerate}
\usepackage{dsfont}
\usepackage{slashed}
\usepackage{float}
\usepackage[titletoc]{appendix}
\usepackage{ulem}

\usepackage{verbatim}

\usepackage{tikz,tikz-3dplot}
\tdplotsetmaincoords{80}{45}

\tikzset{surface/.style={draw=black, fill=white, fill opacity=.6}}

\usepackage{tikz}
\usetikzlibrary{decorations.pathmorphing,patterns}

\newcommand{\be}{\begin{equation}}
\newcommand{\ee}{\end{equation}}
\newcommand{\bea}{\begin{eqnarray}}
\newcommand{\eea}{\end{eqnarray}}

\newcommand{\ben}{\begin{eqnarray}}
\newcommand{\een}{\end{eqnarray}}

\newcommand{\arctanh}{{\rm arctanh}}

\title{Geometric Josephson junction}

\author[a,b]{Fabiano F. Santos}
\author[b]{Henrique Boschi-Filho}

\affiliation[a]{Departamento de Física, Universidade Federal do Maranhão, Campus Universitario do Bacanga, São Luís (MA), 65080-805, Brazil.}
\affiliation[b]{Instituto de Física, Universidade Federal do Rio de Janeiro, Rio de Janeiro, RJ, 21941-909, Brazil}% CEP atualizado
\emailAdd{fabiano.ffs23@gmail.com}
\emailAdd{hboschi@gmail.com}

\abstract{In this work, we present a gravitational dual to a constriction Josephson junction constructed from the AdS/BCFT correspondence. On the gravity side, we consider a planar AdS-Schwarzschild black hole. Our junction is connected by the boundary $\partial\Omega$ with tension $\Sigma$ on the boundary CFT. This approach lead us to analytical solutions rather than usual numerical methods. Our computations on the gravity side reproduce the standard relation between the current across the junction and the phase difference of the condensate controlled by the tension $\Sigma$. We also study the maximum current's dependence on the junction's tension and size and reproduce familiar results.}  

\dedicated{This work is dedicated to the memory of Roberto Nicolsky.}

\begin{document}
	\maketitle
	\newcommand{\limit}[3]
	{\ensuremath{\lim_{#1 \rightarrow #2} #3}}

%\tableofcontents

\section{Introduction}

\begin{figure}[!ht]
\begin{center}
\includegraphics[width=13cm]{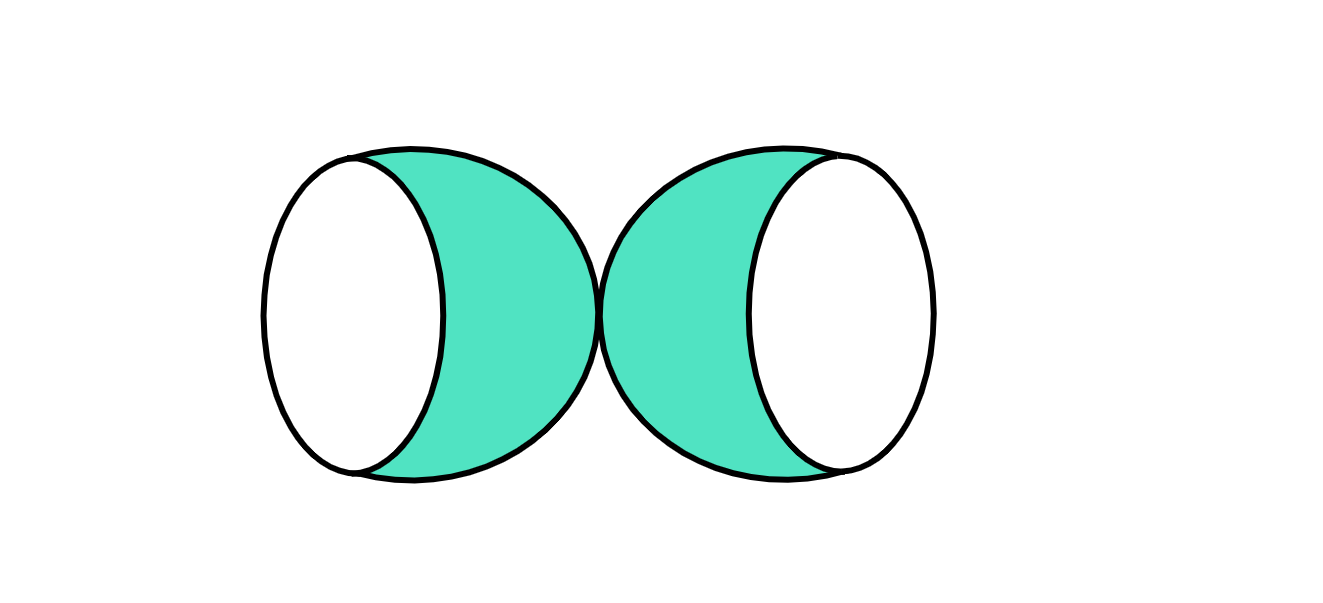}
\caption{Representation of a constriction Josephson junction by gluing together two superconductors of the same material.}
\label{JJconstriction}
\end{center}
\end{figure}

Josephson junctions (JJ) were proposed long ago \cite{Josephson:1962zz} and since then they have been used in many different applications from SQUIDs to qubits (see. {\sl e.g.} \cite{Barone1982, Likharev1986, Tinkham1996, Hays2021}). These junctions are formed by superconducting devices using the tunnel effect, known in this case as the Josephson effect. It allows one to measure quantum behavior in macroscopic settings and are characterized by weak links connecting superconducting electrodes, as in the case of superconductor-insulator-superconductor (SIS), and superconductor-normal-superconductor (SNS) junctions. It is also possible to find the Josephson effect in a single superconductor with a constriction (see Fig. \ref{JJconstriction}) (for a review see, {\sl e.g.} \cite{NicolskyPRB1990, Newrock2000}). This is the case we are interested in this work. Usually, these junctions are described by standard condensed matter theory \cite{Barone1982, Likharev1986, Tinkham1996, Hays2021, NicolskyPRB1990, Newrock2000}, but recently these systems were also studied employing modern AdS/CFT techniques.

Anti-de-Sitter/Conformal Field Theory (AdS/CFT) correspondence \cite{Maldacena:1997re, Gubser:1998bc, Witten:1998qj} relates string theory on asymptotically AdS spacetimes to conformal field theories (CFT) on its boundary.  Since these two theories are defined in spacetimes with different dimensionalities, this duality is usually called holography. The main advantage of this equivalence is that it describes a strongly coupled field theory in terms of a weak coupled gravitational one \cite{Baggioli:2016rdj,Son:2002sd}. In this framework, charged black holes are dual to states of quantum matter at finite temperature and density. 

This duality provides a wide range of possible relations between gravity and, for instance, hydrodynamics
\cite{Policastro:2002se,  Herzog:2002fn, Policastro:2002tn, Kovtun:2003wp}, the quark-gluon plasma 
\cite{Baier:2007ix, Casalderrey-Solana:2011dxg, Herzog:2006gh, Gubser:2006bz}, strong interactions \cite{Boschi-Filho:2006jiv,Mateos:2007ay,BehnamPourhassan:2022rkg},  and condensed matter systems 
\cite{Santos:2023flb,  Hartnoll:2009sz, Hartnoll:2008hs, Hartnoll:2007ai, Hartnoll:2007ih, Hartnoll:2007ip, Hartnoll:2008vx, Arean:2010xd, Horowitz:2011dz,Takeuchi:2013kra,Takeuchi:2021kjv, Domokos:2012rj, Fujita:2012fp, Cai:2013sua, Li:2014xia, Wang:2012yj, Hu:2015dnl, Hovhannisyan:2022gkp, Liu:2022bdu, Melnikov:2012tb}. 
Phenomena like the Hall \cite{Fujita:2012fp, Melnikov:2012tb} and the Nernst \cite{Hartnoll:2007ai, Hartnoll:2007ih, Hartnoll:2007ip, Hartnoll:2008hs} effects have dual gravitational descriptions, as well as superconductivity  \cite{Hartnoll:2008vx}. In this scenario, a charged condensate forms below a critical temperature through a second-order phase transition, and the conductivity (DC) becomes infinite. 
 A holographic description of the SNS  Josephson junction was proposed in \cite{Horowitz:2011dz}, and the case of SIS in \cite{Wang:2012yj}. 
 Other studies of holographic Josephson junctions were given, for instance, in \cite{Takeuchi:2013kra,Takeuchi:2021kjv,Domokos:2012rj, Cai:2013sua, Li:2014xia,  Hu:2015dnl, Hovhannisyan:2022gkp}.

This work proposes a geometrical building of Josephson junctions with constrictions based on Anti-de-Sitter/Boundary Conformal Field Theory (AdS/BCFT). Our model considers that two holographic superconductors described by BCFTs are glued together, as in Fig.  \ref{JJconstriction}.  

%%%%%%%%%%%%%%%%%%%%%%%%%%%%%%%%%%%%%
%%%%%%%  Figura JJ Constriction %%%%%
%%%%%%%%%%%%%%%%%%%%%%%%%%%%%%%%%%%%%

%%%%%%%%%%%%%%%%%%%%%%%%%%%%%%%%
%%%%%%%% End of Figure %%%%%%%%%
%%%%%%%%%%%%%%%%%%%%%%%%%%%%%%%%

The AdS/BCFT correspondence was proposed by Takayanagi \cite{Takayanagi:2011zk, Fujita:2011fp} inspired by \cite{Karch:2000ct, DeWolfe:2001pq, Bak:2003jk, Clark:2004sb, Azeyanagi:2007qj, Cardy:2004hm}
(see also \cite{Magan:2014dwa,Cavalcanti:2018pta,Kanda:2023zse,Geng:2020qvw,Geng:2020fxl,Geng:2021iyq,Santos:2021orr,Geng:2021mic,Sokoliuk:2022llp,Geng:2022slq, Afrasiar:2023nir,Santos:2023mee} for further developments) as an extension of the AdS/CFT correspondence \cite{Maldacena:1997re, Gubser:1998bc, Witten:1998qj}, defining a boundary in the CFT. This extension has found numerous applications, for instance, in transport coefficients in the Hall effect \cite{Fujita:2012fp, Melnikov:2012tb}, fluid/gravity correspondence \cite{Magan:2014dwa}, and the Hawking-Page phase transition \cite{Geng:2020fxl}.

%

%Besides, recent investigations with two holographic CFTs in two dimensions \cite{Geng:2022slq,Geng:2022tfc,Afrasiar:2023nir} have been proposed to obtain Page curves for the Hawking radiation in Jackiw-Teitelboim (JT) braneworld \cite{Jackiw:1984je,Teitelboim:1983ux,Almheiri:2019qdq,Goto:2020wnk}. This setting is performed by two CFT$_{2}$ coupled along a shared interface where the bulk dual geometry has two AdS$_{3}$ spacetimes, truncated by a shared Karch-Randall brane. These geometry setups generalize the entanglement entropy for bipartite states through the island prescription in the compelling lower-dimensional picture. As shown by \cite{Afrasiar:2023nir}, perfect agreement is found in the limit of large brane tension with doubly holographic computations in the bulk geometry. Gravitational duals on AdS$_{D+1}$ with a subcritical Karch-Randall (KR) brane can be used to connect an AdS$_{D}$ gravity located on the brane coupled with a CFT with a non-gravitational bath that shares the same CFT \cite{Randall:1999vf,Karch:2000ct,Brito:2018pwe,Santos:2023eqp,Almheiri:2019psf,Penington:2019npb,Almheiri:2019hni,Almheiri:2019psy}. 

%%%%%%%%%%%%%%%%%%%%%%%%%%%%%%%%%%%%%%%%%%%%%%%%%%%%%%%%%%%%%%%%%%%%%%%%%%%%%%%
This work is organized as follows: In Sec.~\ref{Hornd}, we present the holographic framework within the  AdS/BCFT correspondence, together with the gauge and a complex scalar fields necessary to construct the Josephson junctions. In  Sec.~\ref{black}, we consider  a planar AdS Schwarzschild black hole ansatz in the probe approximation, the equations of motion for the gauge and scalar fields with the corresponding solutions. In Sec.~\ref{profJose}, we obtain the BCFT profile which allow us to get the Josephson current, with the respective phase in terms of the profile tension. Using holographic techniques, we find the expectation value of the Josephson current and its maximum value both presenting exponential decays. These behaviors are related to previous results in the literature identifying a coherence length inversely proportional to the temperature. In Sec.~\ref{conclu}, we present our last comments and conclusions.

%%%%%%%%%%%%%%%%%%%%%%%%%%%%%%%%%%%%%%%%%%%%%%%%%%%%%%%%%%%%%%%%%%%%%%%%%%%%%%%

\section{Holographic framework}\label{Hornd}
In this section, we construct a holographic model to study the constriction Josephson junction. We start reviewing briefly the AdS/BCFT geometry. Then, we move to the discussion of the bulk field actions and surface terms. 

%\subsection{AdS/BCFT setup}

\begin{figure}[ht!]
\begin{center}
\includegraphics[width=11cm]{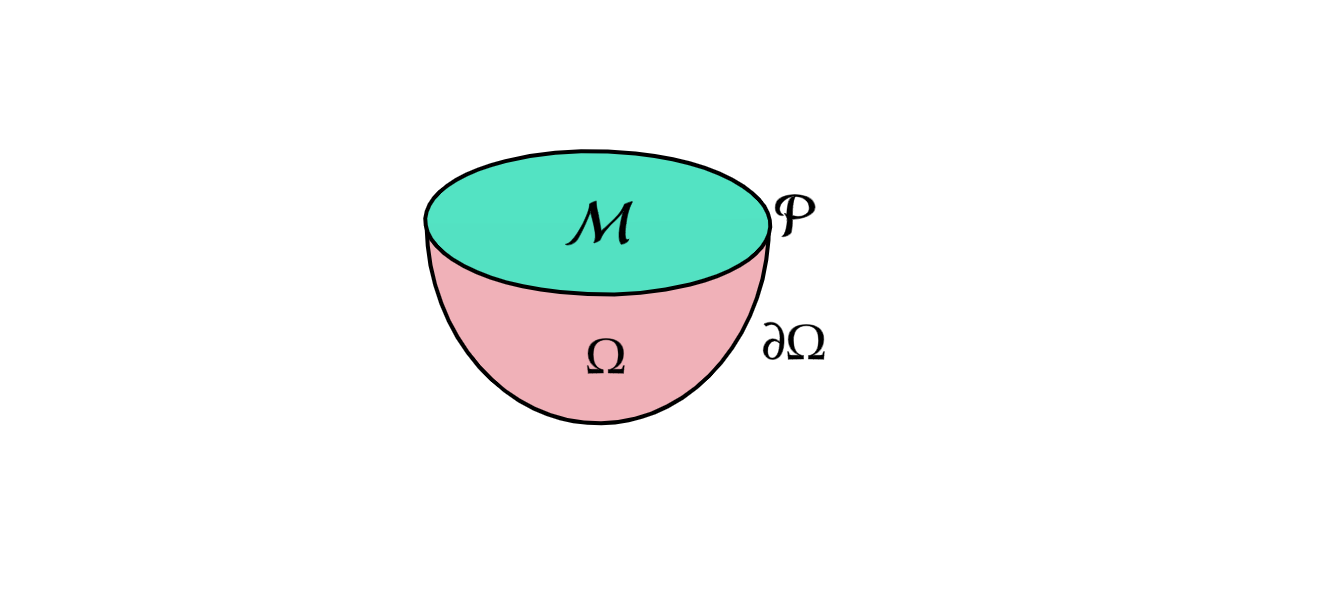}
\caption{AdS/CFT correspondence in the presence of a boundary hypersurface $\partial\Omega$. The CFT is defined on $\mathcal{M}$ with boundary $\mathcal{P}$.} \label{p0}
\label{planohwkhz}
\end{center}
\end{figure}

First of all, we characterize the AdS/BCFT setup for a $D+1$-dimensional asymptotically AdS space $\Omega$ limited by a boundary hypersurface $\partial\Omega$ and a $D$-dimensional manifold $\mathcal{M}$. Note that  $\mathcal{P}=\partial\Omega~\cap~ \mathcal{M}$ is the boundary of the CFT defined on $\mathcal{M}$ (Fig. \ref{p0}).

%%%%%%%%%%%%%%%%%%%%%%%%%%%%%%%%%
%%%%%%%%% Figure ...  %%%%%%%%%%%
%%%%%%%%%%%%%%%%%%%%%%%%%%%%%%%%%

%\vskip0.5cm
%\vskip0.5cm
\begin{figure}[ht!]
\begin{center}
\includegraphics[width=17cm]{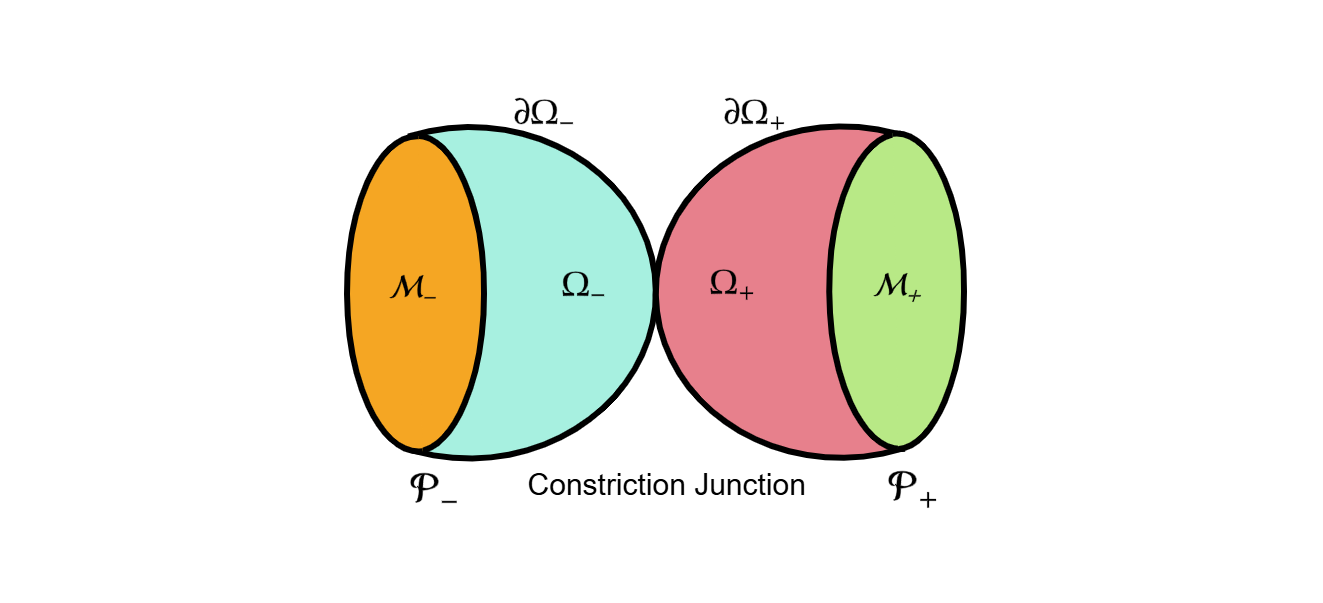}
\end{center}
\caption{This figure represents the bulk manifold $\Omega$ split into the two domains $\Omega_{\pm}$, such that $\Omega=\Omega_{+}\cup\,\Omega_{-}$. In the AdS/BCFT correspondence,  $\Omega_{\pm}$ are asymptotically AdS spaces limited by $\mathcal{M}_{\pm}\cup~\partial\Omega_{\pm}$ where $\partial\Omega_{\pm}$ are  manifolds that satisfy $\partial\Omega_{\pm}\cap~\mathcal{M}_{\pm}=\mathcal{P}_{\pm}$.}
\label{p00}
\end{figure}

Doubling the geometry described in Fig. \ref{p0}, one can construct a  Josephson junction with a constriction as presented in Fig. \ref{p00}.

%\subsection{Bulk description}

In our holographic Josephson junction, two planar AdS$_{4}$  Schwarzschild black holes in bulk \cite{Horowitz:2011dz} theory will be connected by a dual AdS$_{3}$ profile \cite{Takayanagi:2011zk,Fujita:2011fp,Santos:2021orr}, which is described through a CFT$_{3}$ in a thermal state residing in an eternal black hole background.

The geometric profile of the junctions is drawn in Fig. \ref{p00}, where the junction is formed in the BCFT, i.e.,  on the boundary $\partial\Omega=\partial\Omega_{+}\cup~\partial\Omega_{-}$. The action that describes the geometry profile of these junctions is given by
\begin{eqnarray}
\label{SProfile}
S_{\rm Profile}=S^{\Omega}
+S^{\Omega}_{\rm mat}+S^{\partial\Omega}+S^{\partial\Omega}_{\rm mat}+S^{\mathcal{P}}_{\rm ct}, 
\end{eqnarray}
where the bulk action corresponds to the usual Einstein-Hilbert term with negative cosmological constant
\begin{eqnarray}
&&S^{\Omega}=\frac 12 \kappa_{\rm G} \left(\int_{\Omega_{+}}+\int_{\Omega_{-}}\right){d^{D+1}x\sqrt{-g}~(R-2\Lambda)},
\label{SOmega}
\end{eqnarray}
with $\kappa_{\rm G}^{-1}=8\pi\,G$, with $G$ being the gravitational constant, and $g_{\mu\nu}$ the bulk metric defined in the space $\Omega=\Omega_{+}
\cup~\Omega_{-}$. The action 
$S^{\Omega}_{\rm mat}$ describes a perfect fluid while  
\begin{eqnarray} 
&&S^{\partial\Omega}=\kappa_{\rm G} \left(\int_{\partial\Omega_{+}}+\int_{\partial\Omega_{-}}\right){d^{D}x\sqrt{-h}~(K-\Sigma)},
\label{SdOmega}
\\ 
&&S^{\partial\Omega}_{\rm mat}=\kappa_{\rm G} \left(\int_{\partial\Omega_{+}}+\int_{\partial\Omega_{-}}\right){d^{D}x\sqrt{-h}~\mathcal{L}_{\rm mat}}
\label{SdOmegamat}
\end{eqnarray}
are surface terms defined on the boundary $\partial\Omega$, with induced metric $h_{\mu\nu}$, tension $\Sigma$, extrinsic curvature $K_{\mu\nu}=h^{\phantom{\mu}\beta}_{\mu}\nabla_{\beta}n_{\nu}$, with trace $K=h^{\mu\nu}K_{\mu\nu}$, and an outward pointing unit normal vector $n^{\mu}$.  The matter Lagrangian $\mathcal{L}_{\rm mat}$ is  needed to provide an asymptotic AdS spacetime. The action $S^{\mathcal{P}}_{\rm ct}$  contains possible counter-  and contact terms, localized on $\mathcal{P}$:
\begin{eqnarray} 
S^{\mathcal{P}}_{\rm ct} =\left(\int_{\mathcal{P}_{+}}+\int_{\mathcal{P}_{-}}\right){d^{D}x\sqrt{-h}~\mathcal{L}_{\rm ct}}.
\label{SPct}
\end{eqnarray}

In order to describe the  Josephson  junctions we consider a complex scalar field  $\Psi=\tilde{\Psi}\,e^{i\theta}$ where $\tilde{\Psi}$ is a fundamental scalar field, and $\theta$ is the Stückelberg field \cite{Liu:2022bdu}, coupled to gauge fields $A_\mu$ through the Lagrangian:
\begin{eqnarray}
&&{\cal L}_{\rm FF}=-\dfrac{\kappa}{4q^{2}} F^{\mu \nu} F_{\mu \nu}-(\partial_\mu\Psi-qA_\mu)^2-\frac{1}{2}m^2\Psi^2, \label{L3} 
\end{eqnarray}
where $F=dA$ and $m$ is the mass of the scalar field $\Psi$. We will restrict our analysis to the probe approximation scenario. 

 Let us then consider  the following total action
\begin{eqnarray}\label{HF1}
S=S_{\rm Profile} +S^{\Omega}_{FF}
\end{eqnarray}
where $S_{\rm Profile}$ is defined by Eq. \eqref{SProfile} supplemented by Eqs. \eqref{SOmega}-\eqref{SPct} 
and  
\begin{eqnarray}
&&S^{\Omega}_{FF}=\frac{1}{2}\kappa_{\rm G} \left(\int_{\Omega_{+}}+\int_{\Omega_{-}}\right){d^{D+1}\sqrt{-g}\left(-\dfrac{1}{4q^{2}} F^{\mu \nu} F_{\mu \nu}-(\partial_\mu\Psi-qA_\mu)^2-\frac{1}{2}m^2\Psi^2\right)}.\nonumber \\
&& 
\end{eqnarray}
In our setup, we perform a rescale  $\Psi=\tilde{\Psi}/q$, $A=\tilde{A}/q$ and consider $q\to\infty$, keeping $\tilde{\Psi}$ and $\tilde{A}$ fixed. Such considerations are important in neglecting the backreaction of these fields on the metric. 
From the action (\ref{HF1}), we derive the gravitational equations of motion calculating $\delta S^{\Omega}$ and assuming $S^{\Omega}_{\rm mat}=$~constant so that $\delta\,S^{\Omega}_{\rm mat}=~0$, 
\begin{equation}
(G_{\mu\nu}+\Lambda g_{\mu\nu})_{\Omega}=0,
%\quad \Omega=\Omega_{+}\cup~\Omega_{-}
\label{H2}
\end{equation}
%\sigma=\mu and \nu=\rho
and  impose \cite{Santos:2021orr,Santos:2023flb}:
\begin{eqnarray}
K_{\mu\nu}-h_{\mu\nu}(K-\Sigma)=\kappa \,{\cal S}^{\partial\Omega}_{\mu\nu}\,,\label{H7}
\end{eqnarray}
where we defined: 
\begin{eqnarray}
{\cal S}^{\partial\Omega}_{\mu\nu}=-\frac{2}{\sqrt{-h}}\frac{\delta S^{\partial\Omega}_{\rm mat}}{\delta h^{\mu\nu}}\,.\label{H9} 
\end{eqnarray}
Considering the matter stress-energy tensor $S_{\rm mat}$ on $\partial\Omega$ as a constant, which implies  ${\cal S}^{\partial\Omega}_{\alpha\beta}=0$, we can write
\begin{eqnarray}
[K_{\mu\nu}-h_{\mu\nu}(K-\Sigma)]_{\partial\Omega}=0.
%\quad \partial\Omega=\partial\Omega_{+}\cup~\partial\Omega_{-}.
\label{H10}
\end{eqnarray}
This is the dynamical equation for the induced metric $h_{\mu\nu}$, that is, for the profile of the hypersurface $\partial\Omega$. In the next section, we introduce an ansatz for the background solution as a planar AdS black hole so that we will be able to find an explicity solution for the profile in Sec. \ref{profJose}. 

%%%%%%%%%%%%%%%%%%%%%%%%%%%%%%%%%%%%%%%%%%%%%%%%%%%%%%%%%%%%%%%%%%%%%%%%%%%%%%%%%%%%%%%%%
\section{AdS planar Schwarzschild black hole}\label{black}
%%%%%%%%%%%%%%%%%%%%%%%%%%%%%%%%%%%%%%%%%%%%%%%%%%%%%%%%%%%%%%%%%%%%%%%%%%%%%%%%%%%%%%%%%
 In this section, we present a fixed metric background AdS$_4$ planar Schwarzschild black hole
 \begin{equation}
 ds^{2}=-f(u)dt^2+\frac{du^2}{f(u)}+u^2(dx^2+dy^2). 
 \label{BHmetric}
 \end{equation}
 which is a solution for the equation (\ref{H2}) in the probe approximation with 
\begin{eqnarray}
f(u)=\frac{u^{2}}{L^{2}}\left(1-\frac{u^{3}_{h}}{u^3}\right),\label{sol1}
\end{eqnarray}
where $u_{h}$ is the black hole horizon. Note that the coordinate $u$ is defined in the interval $u_h \le u < \infty$ which characterizes the exterior solution of the black hole $(f(u)>0)$, and the boundary $\cal M$ of Fig.  \ref{p0} is located at $u\to \infty$ (the same happens for the boundaries $\cal M_\pm$ of Fig. \ref{p00}. For simplicity, we are going to describe just one solution whenever possible). The solution (\ref{sol1}) resembles the case of \cite{Horowitz:2011dz}. The temperature of this black hole can be written as
\begin{eqnarray}
T=\frac{3u_{h}}{4\pi\,L^2}.\label{sol3}
\end{eqnarray}
Now, we will create the junction along the $y$ direction. We consider solutions of the form
\begin{eqnarray}
\tilde{A}=A_{t}dt+A_{u}du+A_{y}dy,\label{sol4}
\end{eqnarray}
where $A_{t}$, $A_{u}$ and $A_{y}$ are all real functions of $u$ and $y$, as well as $|\Psi|$, defined in Eq. \eqref{L3}. Then, we  define gauge-invariant fields $M=A-d\theta$ and derive the  equations of motion 
\begin{eqnarray}
&\partial^{2}_{u}|\Psi|+\displaystyle{\frac{1}{u^2f}}\partial^{2}_{y}|\Psi|+\left(\frac{f^{'}}{f}+\frac{2}{u}\right)\partial_{u}|\Psi|+\frac{1}{f}\left(\frac{M^{2}_{t}}{f}-fM^{2}_{u}-\frac{M^{2}_{y}}{u^2}-m^2\right)|\Psi|=0,\;\;\;\;\;\label{a}\\
&\displaystyle{\partial^{2}_{u}M_{t}+\frac{1}{u^{2}f}\partial^{2}_{y}M_{t}+\frac{2}{u}\partial_{u}M_{t}-\frac{2|\Psi|^{2}}{f}M_{t}=0,}\label{b}\\
&
\partial^{2}_{u}M_{u}-\partial_u\partial_xM_y-2u^2|\Psi|^2M_u=0,\label{c}\\
&\partial^{2}_{u}M_{y}-\partial_u\partial_y\,M_u+
\displaystyle{\frac{f^{'}}{f}}(\partial_uM_y-\partial_yM_u)-\frac{2|\Psi|^2}{f}M_y=0,\label{d}\\
&
\partial_uM_u+\displaystyle{\frac{1}{u^2f}\partial_yM_y+
\frac{2}{|\Psi|}}\left(M_u\partial_u|\Psi|+\frac{M_y}{u^2f}\partial_y|\Psi|\right)+\left(\frac{f^{'}}{f}+\frac{2}{u}\right)M_u=0.\label{e}
\end{eqnarray}
The equations \eqref{a}-\eqref{d} are of second-order and so describe the dynamics of the problem, while the last one (\ref{e}) is a first-order constraint coming from the current conservation of  Maxwell's equations. Therefore, considering  boundary conditions at $u\to\infty$, one sees that the scalar field has an asymptotic behavior like:
\begin{eqnarray}
|\Psi|=\frac{\Psi^{(1)}(y)}{u}+\frac{\Psi^{(2)}(y)}{u^{2}}+\mathcal{O}\left(\frac{1}{u^{3}}\right), 
\end{eqnarray}
where for simplicity  we consider  $L=1$ and $m=-2$ in this expansion. As discussed in  \cite{Horowitz:2011dz}, both $\Psi^{(1)}$ and $\Psi^{(2)}$ are normalizable, leaving some freedom on these functions. Choosing $\Psi^{(1)}=0$, and using  gauge/gravity duality implies that $\Psi^{(2)}$ is the expectation value of a dimension two operator in the boundary field theory 
\begin{equation}
\langle\mathcal{O}\rangle=\Psi^{(2)}(y), 
\label{psi2}
\end{equation}
 interpreted as the operator of the superconducting condensate. Then, the asymptotic form of the Maxwell fields near the boundary are
\begin{eqnarray}
&&M_t=\mu-\frac{\rho(y)}{u}+\mathcal{O}\left(\frac{1}{u^{2}}\right),\label{Mt} \\
&&M_u=\mathcal{O}\left(\frac{1}{u^{3}}\right),\label{Mu}\\
&&M_y=\nu(y)+\frac{J}{u}.\label{My}
\end{eqnarray}
The quantities $\mu$, $\rho(y)$, $\nu(y)$, and $J$ are interpreted in the boundary field theory as the chemical potential, charge density,  superfluid velocity, and current, respectively \cite{Hu:2015dnl}.  

\section{Profile description of the Josephson junction}
\label{profJose}

\begin{figure}[!ht]
\begin{center}
\includegraphics[width=\textwidth]{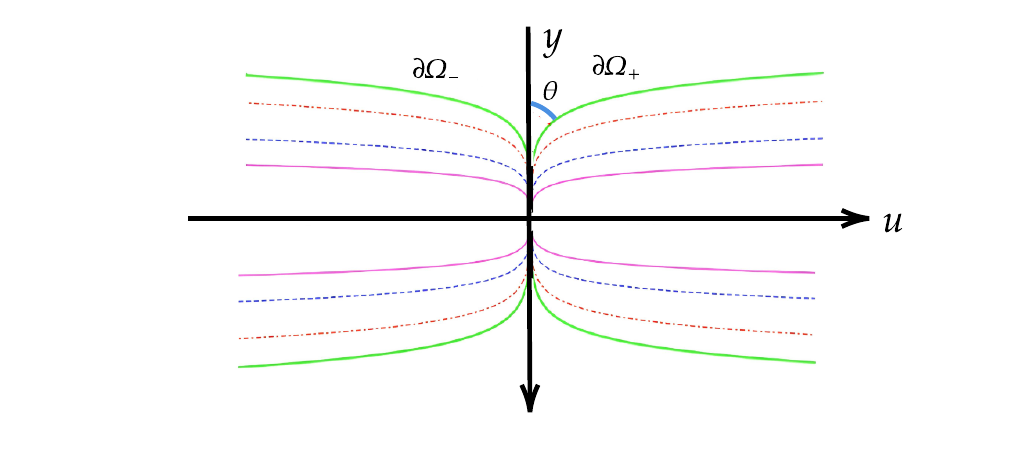}
\caption{This figure presents the junction of two profiles $y(\bar{u})$ where $\bar{u}=u-u_h$ with the two domains $\partial\Omega_{\pm}$ for the values $L=1$ with $\Sigma=0.22$ ({\sl solid}), $\Sigma=0.24$ ({\sl dashed}), $\Sigma=0.26$ ({\sl dot dashed}), and $\Sigma=0.28$ ({\sl thick}).}
\label{figpro}
\end{center}
\end{figure}

From the planar AdS$_4$ black hole metric (\ref{BHmetric}), one can find the normal vectors of the surface $\partial\Omega$ 
\begin{eqnarray}
(n^{t},n^{x},n^{y},n^{u})=\left(0,0,\frac{1}{ug(u)},-\frac{uf(u)y{'}(u)}{g(u)}\right),
\label{4}
\end{eqnarray}
where $g^{2}(z)=1+u^2f(u)y'^{2}(u)$ with  $y{'}={dy}/{du}$ and the induced metric on $\partial\Omega$ is an AdS$_3$ space given by
\begin{eqnarray}
 ds^{2}_{ind}=-f(u)dt^2+\frac{g^2(u)du^2}{f(u)}+u^2dx^2. 
 \label{metricind}
\end{eqnarray}
Now, solving the dynamical equation (\ref{H10}) for the induced metric on the boundary $\partial\Omega$, we find 
\begin{eqnarray}
y{'}_{\pm}(u)=\pm\frac{\Sigma}{\sqrt{4+\Sigma^{2}u^2f(u)}}.\label{prof}
\end{eqnarray}  
It is possible to integrate this expression in terms of elliptic integrals. Considering the approximation $f(u)\approx u^2/L^2$, valid for large $u$, and discarding terms of order ${\cal O}\left(\frac{1}{u^5}
 \right)$, one finds 
\begin{eqnarray}
 u \approx \frac{1}{|y|},  \qquad\qquad 
 (u\gg u_h). 
\label{udey}
\end{eqnarray} 
The behavior of the above solutions is presented in Fig. \ref{figpro}. Then, the constriction Josephson junctions are formed by gluing together two of these profiles. 

The phase difference between the two superconductors is given by gauge invariance
\begin{eqnarray}
\gamma=\Delta\theta-\int{A_{y}},
\end{eqnarray}
where the integral crosses the constriction. In our case, we can write
\begin{eqnarray}
\gamma=-\frac{1}{J}\left(\int_{-\infty}^{-u_h}+\int_{u_h}^{\infty}\right)
\nu(y)\, dy
\label{phase}
\end{eqnarray} 
%
%The equations (\ref{a})-(\ref{e}) are invariant under the following scaling symmetry 
%\begin{eqnarray}
%u\to\,au,\quad(t,x,y)\to(t,x,y)/a,\quad\,M_{u}\to\,M/a,\quad(M_{t},M_{y})\to\,a(M_{t},M_{y}).
%\end{eqnarray}
%For the case of homogeneous superconductors, the scale invariant temperature is $T/\mu$, so changing $\mu$ is equivalent changing the temperature $T$, so $T$ must be a constant. In Ref. \cite{Horowitz:2011dz}, it was used a non-homogeneous chemical potential $\mu(x)$ to control the temperature of the superconductor and define the SNS JJ profile including its width~$\ell$. In our case, the BCFT profile  $y(u)$ plays the hole of $\mu(x)$. 
%Actually, if one imposes a boundary condition on the gauge field $M_t$, Eq. \eqref{Mt}, at some value of the holographic coordinate $u=u_0$, one would find a relation between the chemical potential $\mu(u)$, the charge density $\rho(y)$ and the holographic coordinate $u$ itself as $\mu(u)=\rho(y)/u$. 

\begin{figure}[!ht]
\begin{center}
\includegraphics[scale=0.5]{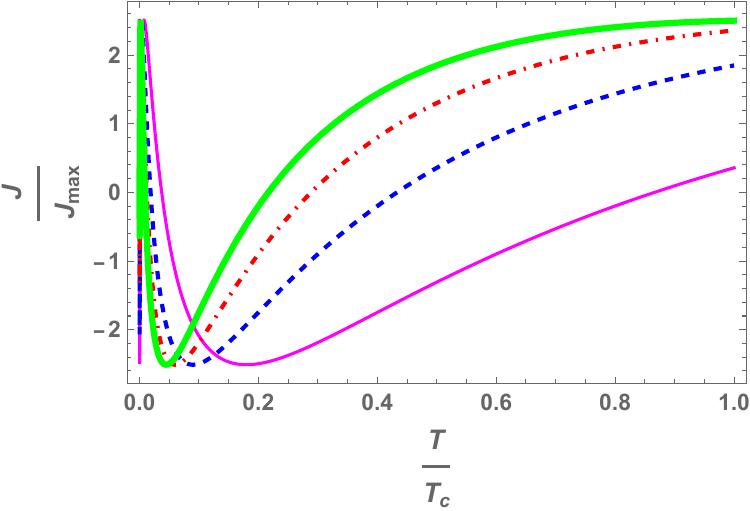}
\includegraphics[scale=0.5]{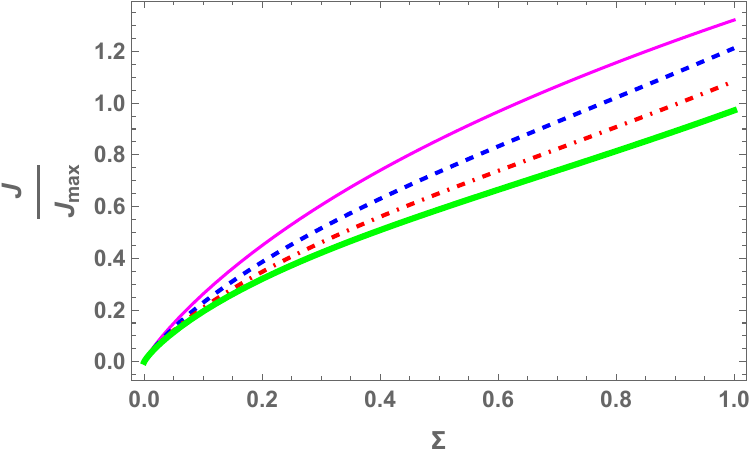}
\caption{This figure shows the behavior of the ratio $J/J_{max}$ given by Eq. (\ref{currentjsp}) with the phase \eqref{phase2} 
against $T/T_c$ for various values of $\Sigma$ (left panel) and against $\Sigma$ for various values of $T/T_c$ (right panel). Left panel: $\Sigma=0.1$ ({\sl solid pink}), $\Sigma=0.2$ ({\sl  dashed blue}), $\Sigma=0.3$ ({\sl dot dashed red})  and $\Sigma=0.4$ ({\sl thick green}). Right panel:  $T/T_c=0.1$ ({\sl  solid pink}), $T/T_c=0.2$ ({\sl dashed blue}), $T/T_c=0.3$ ({\sl dot dashed red})  and $T/T_c=0.4$ ({\sl thick green}).}
\label{holographicJJ}
\end{center}
\end{figure}

 Using Dirichlet boundary conditions in gauge-invariant fields $n^{\mu}M_{\mu}|_{\partial\Omega}=0$, one finds 
\begin{eqnarray}
\nu(y)=-\,\frac{J}{u}, 
\label{nu(y)}
\end{eqnarray}
where $y=\int{du\,y{'}(u)}$ comes from the solutions \eqref{prof}. So, the Josephson's current reads 
\begin{eqnarray}
J=J_{max}\sin(\gamma),\label{currentjsp}
\end{eqnarray}
and exists even in the absence of an external applied voltage. 
% The holographic  Josephson current flows from the disk $\tau^2+y^2\leq\,r_{D}^2$ where $\tau$ is the Euclidean time (see Fig. \ref{DISK}), on the surface $\mathcal{M}$ to the boundary at $\partial\Omega$ \cite{Takayanagi:2011zk,Fujita:2011fp}. 
From Eqs. \eqref{phase} and \eqref{nu(y)}, one has 
\begin{eqnarray}
\gamma&=&\int^{-u_h}_{-\infty}{du\frac{y{'}(u)}{u}}-\int^{\infty}_{u_h}{du\frac{y{'}(u)}{u}}\nonumber \\ 
      &=&-\Sigma\,\int^{-u_h}_{-\infty}{\frac{du}{u\sqrt{4+(\Sigma\,u)^2f(u) }}}+\Sigma\,\int^{\infty}_{u_h}{\frac{du}{u\sqrt{4+(\Sigma\,u)^2f(u) }}}.\label{phase1}
\end{eqnarray} 
Performing these integrals, one finds the phase
\begin{eqnarray}
\gamma=\frac{\Sigma}{2}\, 
\arctanh
\left[
\sqrt{4+\frac{\Sigma^2u^{2}_{h}}{L^2}}\;\right].
\label{phase2}
\end{eqnarray}
The behavior of the Josephson current (\ref{currentjsp}) with phase given by \eqref{phase2} 
as a function of the ratio $T/T_c$ and the tension $\Sigma$ is shown in the two panels of Fig. \ref{holographicJJ}. 
In particular, note that $J$ increases with growing $\Sigma$. 

%\begin{figure}[!ht]
%\begin{center}
%\includegraphics[width=\textwidth]{JD.pdf}
%\caption{The holographic dual of a disk of radius $r_D$.}
%\label{DISK}
%\end{center}
%\end{figure}

To extract the expectation values of the relevant operators of the Josephson junctions here, we follow the general approach of Refs.  \cite{Baggioli:2016rdj,Son:2002sd} where a field $\varphi$ living in bulk can be related to an operator of the CFT with the same quantum numbers, and their coupling shows up through a boundary term.  From the point of view of the AdS/CFT correspondence, a deformation due to the source $\varphi_0$, is given by:
\begin{eqnarray}
\mathcal{L}_{CFT}+\int{d^{d}x\varphi_0\mathcal{O}}
\end{eqnarray}
and through Quantum Field Theory (QFT) \cite{Baggioli:2016rdj} the functional generator of correlation functions $W(\varphi_0)$ can be written as
\begin{eqnarray}
e^{W(\varphi_0)}\langle\,e^{\int{\varphi_0\mathcal{O}}}\rangle_{QFT}. 
\end{eqnarray}
Applying the methodology of \cite{Baggioli:2016rdj,Son:2002sd}, we can derive the correlation functions of an operator $\mathcal{O}$, taking functional derivatives of the generating functional with respect of the source term $\varphi_0$:
\begin{eqnarray}
(\langle \mathcal{O}{...}\mathcal{O} \rangle)_{1...n}=\frac{\delta^{n}W}{\delta\varphi^{n}_{0}}\mid_{\varphi_{0}=0}. 
\end{eqnarray}
 This process allows us to systematically explore the boundary theory's response to perturbations in the bulk. 

  \begin{figure}[t]
\begin{center}
\includegraphics[scale=0.5]{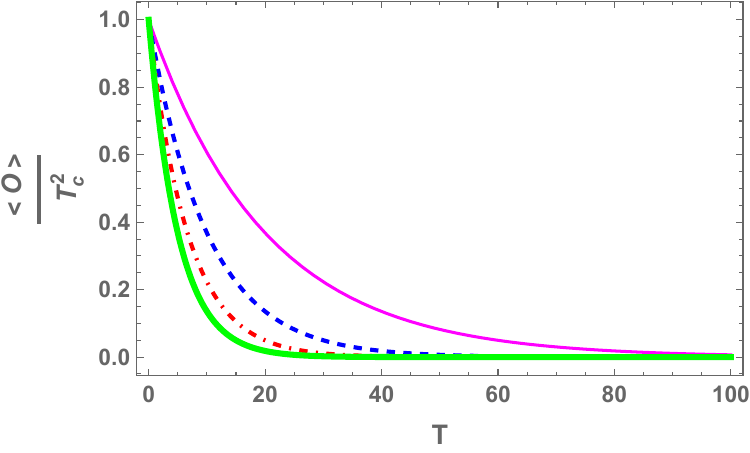}
\includegraphics[scale=0.5]{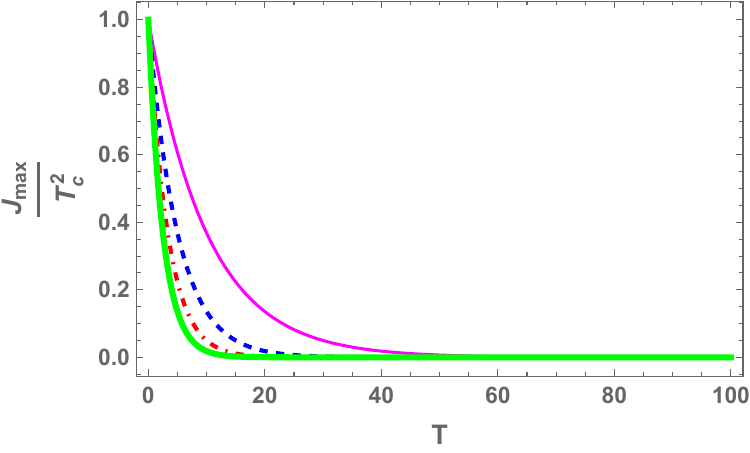}
\caption{This figure shows the behavior of the equations (\ref{operator0}) and (\ref{operator2}) for the values $A_{0}=A_{1}=1$ with $\Sigma=0.1$ ({\sl  solid pink}), $\Sigma=0.2$ ({\sl dot dashed blue}), $\Sigma=0.3$ ({\sl dashed red})  and $\Sigma=0.4$ ({\sl thick green}), respectively.}\label{magnetichgxx}
\end{center}
\end{figure}

 Following the above prescriptions, the expectation value of the superconducting condensate $\langle\mathcal{O}\rangle=\Psi^{(2)}(y)$, Eq. \eqref{psi2},  in the boundary field theory  is given by
\begin{eqnarray}
\langle\mathcal{O}\rangle_{y=0,J=0}={C_{1}}e^{-{\displaystyle{ u_h\,\Sigma/2L}}},
\label{operator0}
\end{eqnarray}
 where $C_1$ is an integration constant. This result can be compared with \cite{Horowitz:2011dz} 
\begin{eqnarray}
\langle\mathcal{O}\rangle_{y=0,J=0}={A_{1}}T^{2}_{c}{\displaystyle{e^{-{\ell}/{2\xi}}}},
\label{operator1}
\end{eqnarray}
if we identify $C_1=A_1T_c^2$, where $T_c$ is the critical temperature. The comparison of Eqs. \eqref{operator0} and 
\eqref{operator1} suggests 
 that $\Sigma$ plays the role of the width $\ell$ of the junction and the ratio $L/u_h$ as the coherence length $\xi$. Since $u_h\propto T$, from Eq. \eqref{sol3}, we expect that the coherence length in our model grows with decreasing temperatures. 

One can also find the behavior of the maximum value of the Josephson current $J_{max}$ against the tension $\Sigma$ and the coherence length $\xi$:
\begin{eqnarray}
\frac{J_{max}}{T^{2}_{c}}={A_{0}}\, {\displaystyle{e^{{-\Sigma}/{\xi}}}}.
\label{operator2}
\end{eqnarray}
The behavior of Eqs. \eqref{operator0} and \eqref{operator2} against the temperature are shown in Fig. \ref{magnetichgxx}, with clear exponential decays in both cases.

 %\newpage

\section{Comments and Conclusions}\label{conclu}

In this work we proposed a holographic model for constriction Josephson junctions based on the AdS/BCFT correspondence. This work was greatly inspired by the holographic superconductor  
\cite{Hartnoll:2008vx} and the SNS and SIS holographic Josephson junctions of Refs. \cite{Horowitz:2011dz, Wang:2012yj}. Our main contribution here is to use the AdS/BCFT instead of the `pure' AdS/CFT correspondence. The AdS/BCFT approach lead us to a profile $y(u)$ which can be solved analytically and allowed  analytical solutions for the phase $\gamma$ of the superconducting current $J=J_{max}\sin(\gamma)$. These led to analytical expressions for the exponential decays of the  condensate $\langle\mathcal{O}\rangle=\Psi^{(2)}(y)\propto \exp\{-\Sigma/2\xi\}$ and of the maximum value of the current $J_{max}\propto \exp\{-\Sigma/\xi\}$, where the coherence length $\xi$ is identified as $L/u_h\sim L/T$. This suggests a decreasing coherence length with increasing temperature. 

The comparison of our results with the ones from Ref. \cite{Horowitz:2011dz} also suggests that the role played by the width $\ell$ of the SNS JJ is perfomed by the tension $\Sigma$ in our constriction JJ. 

It is important to mention that our analytical solutions come from a four dimensional planar AdS Schwarzschild black hole, related to a three-dimensional boundary theory, characterized by the profile function $y(u)$. Higher dimensional models are usually more involved and might require additional numerical treatments. 

We also considered the chemical potential $\mu$ to be constant here and the possibility of a spatial dependence on it might improve our model. Another upgrade that one can look for is the inclusion of a time dependence on a voltage $V$ applied across the junction to the gauge potential $A_t$, since the Josephson effect predicts that the phase difference satisfies
\begin{eqnarray}
\frac{d\gamma}{dt}=qV, 
\end{eqnarray}
which implies an alternating current. In this case, one could change the phase of the charged scalar $\Psi$ by $qVt$. However, time dependence makes it more difficult to construct appropriate gravitational solutions.

It would also be interesting to consider the inclusion of magnetic fields in our set up. Eventually, this might be compared with the results of Ref. 
\cite{Hovhannisyan:2022gkp}.

Further, we intend to extend these results by considering modified theories of gravity rather than standard Einstein's General Relativity, for instance, introducing a Horndeski scalar field possibly representing a Josephson junction with a topological defect.

\acknowledgments

This work was supported by Coordenação de Aperfeiçoamento de Pessoal de Nível Superior (CAPES) under finance code 001.  Conselho Nacional de Desenvolvimento Cient\'{\i}fico e Tecnol\'{o}gico (CNPq) partially funds H.B.-F. under grants $\#$s~311079/2019-9 and 310346/2023-1.

\end{document}